\documentclass[journal]{IEEEtran}
\ifCLASSINFOpdf
  \usepackage[pdftex]{graphicx}
\else
\fi
\usepackage{amsmath,graphicx,epstopdf}
\usepackage{multirow}
\usepackage{url}
\usepackage{dingbat}
\usepackage{hyperref}
\usepackage{enumitem}

\begin{document}

\onecolumn % make sure you keep this coverpage as one column. In this template, we force the coverpage to be one column with this command and then switch to double column for the remaining of the paper with the \doublecolumn command.
\begin{description}[style=multiline,leftmargin=3cm]

\item[\textbf{Citation}]{Yuting~Hu, Zhiling~Long, Anirudha~Sundaresan, Motaz~Alfarraj, Ghassan AlRegib, Sungmee Park, and Sundaresan~Jayaraman, ``Fabric Surface Characterization: Assessment of Deep Learning-based Texture Representations Using a Challenging Dataset,'' The Journal of the Textile Institute, 2020.}

%\item[\textbf{DOI}]{\url{https://doi.org/10.1109/LSP.2018.2825309}}

\item[\textbf{Review}]{Date of acceptance: 12 March 2020}

\item[\textbf{Codes}]{\url{https://github.com/olivesgatech/textureclassification_MULTER_ICIP2019}}% If you do not have data related to this paper, you can remove the data keyword.

\item[\textbf{Dataset}]{\url{https://github.com/olivesgatech/CoMMonS}}% If you do not have data related to this paper, you can remove the data keyword.

\item[\textbf{Bib}] {@article{hu2020deeptexture,\\
  title={Fabric Surface Characterization: Assessment of Deep Learning-based Texture Representations Using a Challenging Dataset},\\
  author={Yuting~Hu, Zhiling~Long, Anirudha~Sundaresan, Motaz~Alfarraj, Ghassan AlRegib, Sungmee Park, and Sundaresan~Jayaraman},\\
  journal={The Journal of The Textile Institute},\\
  year={2020}
}

}

% Preprint sharing policy can vary depending on the publisher. Before posting a paper to arXiv, please specifically check the transaction/convference you are targeting. In some transactions, papers are usually added to arxiv after acceptance. Pubslishers usually allow the authors to share accepted version of their papers but not the final formatted version that is provided by the pubisher. In case of sharing preprints, publishers usually ask to add DOI and citation to the paper along with a copyright notice.

%\item[\textbf{Copyright}]{\textcopyright 2018 IEEE. Personal use of this material is permitted. Permission from IEEE must be obtained for all other uses, in any current or future media, including reprinting/republishing this material for advertising or promotional purposes,
%creating new collective works, for resale or redistribution to servers or lists, or reuse of any copyrighted component
%of this work in other works. }

\item[\textbf{Contact}]{\href{mailto:huyuting@gatech.edu}{huyuting@gatech.edu}  OR \href{mailto:zhiling.long@ece.gatech.edu}{zhiling.long@ece.gatech.edu} OR \href{mailto:alregib@gatech.edu}{alregib@gatech.edu}\\
    \url{http://ghassanalregib.com/} \\ }
\end{description}
%
%%Following command sequence was used to start the paper content from the following page and avoid numbering cover page.
%\thispagestyle{empty}
%\newpage
%\clearpage
%\setcounter{page}{1}

\twocolumn

%\title{Challenging Datasets and CNN-based Texture Attribute Analysis and Assessment}
\title{Fabric Surface Characterization: Assessment of Deep Learning-based Texture Representations Using a Challenging Dataset}
\author{Yuting~Hu,~\IEEEmembership{}
        Zhiling~Long,~\IEEEmembership{}
        Anirudha~Sundaresan,~\IEEEmembership{}% <-this % stops a space
        Motaz~Alfarraj,~\IEEEmembership{}
        Ghassan~AlRegib,~\IEEEmembership{}\\
        Sungmee Park,~\IEEEmembership{}
        Sundaresan~Jayaraman~\IEEEmembership{}

%\address{Georgia Institute of Technology, Atlanta, GA 30332-0250, USA}

% \thanks{M. Shell was with the Department
% of Electrical and Computer Engineering, Georgia Institute of Technology, Atlanta,
% GA, 30332 USA e-mail: (see http://www.michaelshell.org/contact.html).}% <-this % stops a space
% \thanks{J. Doe and J. Doe are with Anonymous University.}% <-this % stops a space
% \thanks{Manuscript received April 19, 2005; revised August 26, 2015.}
}


\markboth{The Journal of The Textile Institute,~Vol.~, No.~, December~2019}%
{Shell \MakeLowercase{\textit{et al.}}: Bare Demo of IEEEtran.cls for IEEE Journals}



\maketitle

\begin{abstract}

Tactile sensing or fabric hand plays a critical role in an individual's decision to buy a certain fabric from the range of available fabrics for a desired application. Therefore, textile and clothing manufacturers have long been in search of an objective method for assessing fabric hand, which can then be used to “engineer” fabrics with a desired hand. Recognizing textures and materials in real-world images has played an important role in object recognition and scene understanding. In this paper, we explore how to computationally characterize apparent or latent properties (e.g., surface smoothness) of materials, i.e., computational material surface characterization, which moves a step further beyond material recognition. We formulate the problem as a very fine-grained texture classification problem, and study how deep learning-based texture representation techniques can help tackle the task. We introduce a new, large-scale challenging microscopic material surface dataset (CoMMonS), geared towards an automated fabric quality assessment mechanism in an intelligent manufacturing system. We then conduct a comprehensive evaluation of state-of-the-art deep learning-based methods for texture classification using CoMMonS. Additionally, we propose a multi-level texture encoding and representation network (MuLTER), which simultaneously leverages low- and high-level features to maintain both texture details and spatial information in the texture representation. Our results show that, in comparison with the state-of-the-art deep texture descriptors, MuLTER yields higher accuracy not only on our CoMMonS dataset for material characterization, but also on established datasets such as MINC-2500 and GTOS-mobile for material recognition. Our dataset and source code are published online at \url{https://ghassanalregib.info/software-and-datasets}, which will serve as a benchmark for evaluating deep learning-based techniques for both material surface characterization and, more generally, very fine-grained texture classification.

\end{abstract}


% \fi
\IEEEpeerreviewmaketitle



% \hfill September 13, 2018

% chapter 1:
\section{Role of Texture in Objects and Scenes}
\label{sec:intro}

\IEEEPARstart{I}{n} his seminal paper, Peirce~\cite{peirce193026} proposed the concept of fabric ``handle'' and stated the following: ``\textit{In judging the feel or `handle' of a material, use is made of such sensations as stiffness or limpness, hardness or softness, and roughness or smoothness. It is desirable to devise physical tests that analyse and reflect the sensations felt and assign numerical values to the measurements}.'' He then presented a series of laboratory tests that reflected the sensations, namely, stiffness, hardness and smoothness (stroking a material), and laid the foundation for linking ``objective'' measurements of fabric properties to ``subjective'' assessment of fabric hand or feel experienced by the user.

\begin{figure}[htb]
\centering
{\includegraphics[width=8.5cm]{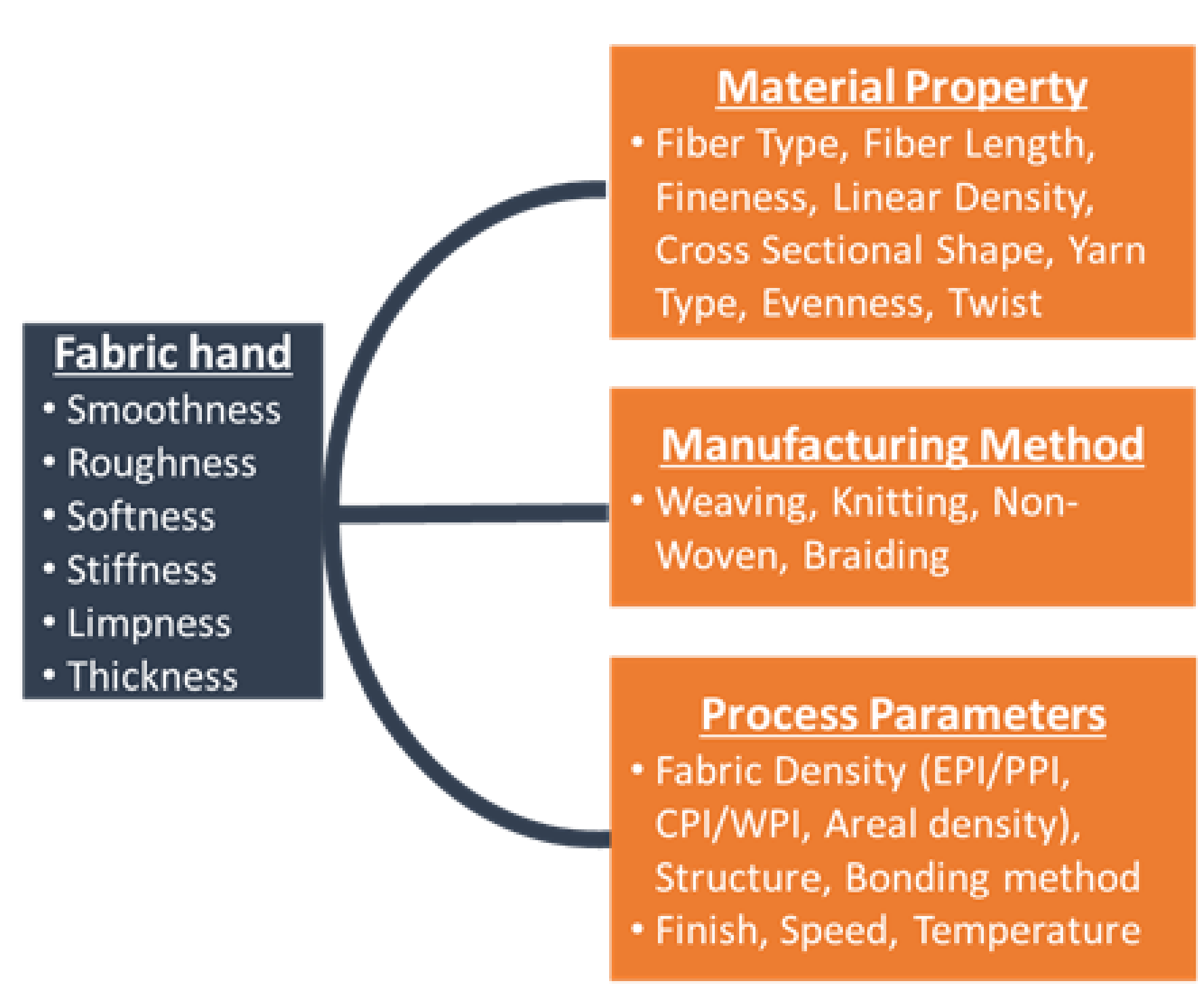}
\caption{Fabric hand: the roles of material, manufacturing and process parameters.}
\label{fig:Materialandparameter}}
\end{figure}

A fabric's subjective properties of stiffness, hardness, smoothness and drape experienced by the user are determined by the intrinsic properties of the materials, (e.g., fibers and yarns), manufacturing methods (e.g., weaving, knitting, braiding, nonwovens) and process parameters (e.g., structure, speeds, finishes). The relationship between these independent parameters - material, manufacturing, and process - and the dependent parameter - fabric hand - is shown in Fig.~\ref{fig:Materialandparameter}.

Tactile sensing or fabric hand plays a critical role in an individual's decision to buy a certain fabric from the range of available fabrics for a desired application. Therefore, textile and clothing manufacturers have long been in search of an objective method for assessing fabric hand, which can then be used to ``engineer'' fabrics~\cite{rajamanickam1998structured} with a desired hand using the independent parameters shown in Fig.~\ref{fig:Materialandparameter}. One of the successful systems in practice has been the Kawabata Evaluation System (KES) developed by Kawabata and Niwa to assess the total hand value of fabrics~\cite{kawabatra1980standardization,kawabata2017objective}. The Fabric Assurance by Simple Testing (FAST) system was developed in the 1980s as a simpler alternative to the Kawabata system and minimized the testing and assessment time~\cite{NGLy1990}. Ciesielska-Wrobel and Langenhove present a review of research developments in the area of fabric hand and discuss the challenges associated with the ``subjective'' ranking of fabric hand by humans, especially when the differences are very little~\cite{ciesielska2012hand}.

In short, there is a critical need for a tool or system that can “objectively” assess fabric hand. This is especially so in today's global market in which the enterprise must fine-tune its processes and leverage its core design and manufacturing competencies to produce the \textit{right} product, with the \textit{right} quality, in the \textit{right} quantity, at the \textit{right} price, and at the right time~\cite{srinivasan1999changing}~(the five \textit{R}s). The integration of information technology in the textile industry has been gaining ground in recent years. The advancements in computing and communication technologies are being harnessed to catalyze the textile industry to create customized high quality products that meet the specific needs of consumers. For instance, advanced computer vision systems have been developed to study the features (e.g., texture) of fabrics to identify and classify defects~\cite{balakrishnan1998fdics}.

In this paper, we propose an innovative concept of using computer vision and image processing to study the visual features of fabrics and develop an objective method to assess fabric handle. Unlike the Kawabata or FAST methods used in the textile industry, there is no need to measure a series of mechanical properties of fabrics in the laboratory to assess fabric hand.

\textbf{Role of Texture in Objects and Scenes}: Real-world images exhibit abundant textural information in objects and scenes. To understand such images, it is very important to be able to characterize the textures. Texture representation aims to extract descriptive features that provide important visual cues or characteristic properties of texture patterns. It has been an active research area for decades. Systematic reviews on widely adopted texture descriptors, both handcrafted and learning-based ones, have been provided in~\cite{cimpoi2015deep,liu2017local,liu2019bow,zheng2016deep}. Distinctive and robust representation of texture is the key for various multimedia applications such as content-based image retrieval (CBIR)~\cite{chun2008content}, face detection/recognition~\cite{zhao2017dynamic}, object detection, image/texture segmentation, dynamic texture/scene recognition~\cite{arashloo2014dynamic,zhao2017dynamic,zhao2019dynamic}, and texture/color style transfer~\cite{song2017sufficient,lo2016example}. Commonly, textures observed in natural scenes are representative of various materials. Thus, an effective texture representation will also be helpful for analyzing and understanding the associated materials. Recently, automated material analysis has attracted increasing interest from researchers in the image processing and computer vision community, for potential applications spanning from scene understanding to robotics and industrial inspection~\cite{zheng2016deep}. For example, if an autonomous vehicle knows what kind of ground terrain it is driving on (i.e., whether the ground surface material is concrete, asphalt, soil, or pebbles), it can adjust itself according to the actual outdoor environment to ensure a successful operation~\cite{xue2017differential}.

Existing research works on automated material analysis mainly focus on material recognition, where the objective is to classify various types of materials into their associated categories. Typically, such classification is coarse-grained, with the materials of interest cover a wide range of generic categories. For example, in~\cite{sharan2009material}, materials considered include fabric, foliage, glass, leather, metal, paper, plastic, stone, water, and wood; while in~\cite{xue2017differential,xue2018deep},  materials being studied are ground surfaces made of asphalt, concrete, pebbles, soil, etc.. On a few occasions, fine-grained material classification has been studied, where the subject materials belong to the same generic category but different sub-categories. As an example, in~\cite{kampouris2016fine}, the generic category of fabric is divided into sub-categories such as cotton, terrycloth, denim, fleece, nylon, polyester, silk, viscose, and wool. Although material recognition is important, we believe material analysis should also address the topic of material surface characterization. This is an analysis that takes a step further beyond merely recognizing the material. It aims to find out more detailed information about a material in terms of certain specific properties of the material. For example, for a robotic arm to catch a glass container, not only does it need to recognize the material being glass (thus being fragile), but it also has to know further the level of fragility of that particular glass container, so that it can apply appropriate level of pressure to the object when catching it. Such problems have rarely been reported in the image processing and computer vision literature, but are of practical significance. Essentially, material characterization can be considered as a very fine-grained classification, where the categories belong to the same material (e.g., glass) but represent different levels in terms of a certain property of the material (e.g. fragility).

In this paper, we study material surface characterization in the context of intelligent manufacturing systems in the textile industry, where an automatic quality evaluation system is in need to perform an objective assessment of manufactured fabric surfaces, viz., fabric hand. Traditionally, the hand assessment is performed by a subject expert who manually touches the fabric. Not only does such a subjective assessment demand skills and experience, but it also has major drawbacks including involving extensive labor, consuming considerable amount of time, and most undesirably, suffering from possible human errors and inconsistency. Therefore, it is desirable to develop an automatic system that can perform an objective fabric surface or hand assessment instead, which is reliable, consistent, and efficient. The objective is to examine the fabrics in terms of the characteristics of interest such as the relative fiber length and smoothness on the fabric surface. Few attempts reported in literature (e.g.,~\cite{ujevic2009analysis}) have tried to characterize fabrics via objective means using chemical, physical, or mechanical measurements, but with limited success. We believe texture image analysis is a feasible approach here, as the target characteristics are visual and tactile properties observed at fabric surfaces which have been demonstrated to be generally correlated with visual properties of texture images~\cite{tamura1978textural,hu2018high}.

Treated as a very fine-grained texture classification, the task of material surface characterization is particularly challenging in two aspects. First, comparing to coarse-grained classification, here the inter-class appearance variations are much smaller, while the intra-class visual variations can sometimes be relevantly significant because of diverse photometric and geometric conditions. Second, some material surface characteristics of interest may be more latent than apparent, not always easily distinguishable in the visual appearance. Therefore, advanced techniques such as deep learning-based texture analysis are indispensable for the material surface characterization task.

Nowadays, because of the record-breaking recognition performance, the convolutional neural network (CNN)~\cite{krizhevsky2012imagenet} has emerged as the new state-of-the-art tool for object recognition and classification. Different from object recognition, texture and material recognition generally is challenging in demanding an orderless representation of micro-structures (i.e., texture encoding). Concatenated global CNN activations with a fully connected layer as a classifier have limitations in meeting the need for a geometry-invariant representation describing feature distributions. To transfer knowledge from object recognition to texture recognition, several CNN-inspired approaches were proposed~\cite{liu2017local,liu2019bow}. As the most representative work among these initial attempts, Cimpoi et al.~\cite{cimpoi2015deep} proposed a learning-based texture representation, FV-CNN, that replaces handcrafted filter banks with pretrained convolution layers for the feature extractor. One shortcoming of the FV-CNN architecture is the separate learning of CNN feature extraction, texture encoding and classifier training, which does not benefit from the labeled data. To jointly learn them together in an end-to-end manner, Zhang et al.~\cite{zhang2017deep} proposed a deep texture encoding network (Deep-TEN), which builds a texture encoding layer incorporating the dictionary learning and feature pooling on top of the CNN architecture. Later, Xue et al.~\cite{xue2018deep} presented a deep encoding pooling network (DEP), which adds local spatial information to Deep-TEN. However, neither Deep-TEN nor DEP fully utilizes CNN features from different layers and resolutions. Therefore, in this paper, we propose a multi-level convolutional neural network that improves over them in this aspect.

As mentioned earlier, our research is targeted towards material surface characterization. To the best of our knowledge, the only deep learning-based material characterization work in the literature was reported by Sun et al.~\cite{sun2018noncontact}, who automatically estimated roughness of milled metal surfaces using a CNN. However, our work is different from theirs in the following aspects. First, our work is based on texture analysis. As we will demonstrate, the algorithm we develop serves as a general texture representation technique, thus applicable to other texture analysis problems. In contrast, the work in~\cite{sun2018noncontact} is purely image-based, not really oriented towards textures. Second, our work provides an end-to-end deep-learning solution, while theirs is not. Third, our work is generally applicable to material surface characterization in terms of all kinds of material properties. However, their solution consists of modules that are specifically designed for metal surface roughness, not generalizable for other materials and properties.

Similar to ImageNet~\cite{russakovsky2015imagenet} for object recognition and natural scene classification, large-scale texture and material image datasets have been created through both Internet-based crowd-source annotation and in-lab controlled acquisition in recent years. Among these datasets, the Flickr material database (FMD)~\cite{sharan2009material} and describable textures dataset (DTD)~\cite{cimpoi2014describing} are moderate-sized datasets for recognizing describable texture attributes in natural images. The materials in context database (MINC)~\cite{bell2015material} is a large-scale, diverse and well-sampled dataset, with 23 categories and 3 million material samples. By including Flickr images and Houzz photos from Internet, the MINC dataset represents a wide range of materials in the real world, facilitating material recognition in unconstrained conditions (i.e. in the wild). Most recently, a large-scale material surface dataset, ground terrain in outdoor scenes (GTOS), with over 30,000 images covering 40 classes, was collected by Xue et al.~\cite{xue2017differential}, geared towards real-world material recognition for autonomous agents. These datasets were all created for material recognition, but not suitable for material surface characterization where very fine-grained classification is performed within one type of material in terms of some certain material characteristics or properties. For this purpose, in this paper, we introduce a large, publicly available dataset named challenging microscopic material surface dataset (CoMMonS). We utilize a powerful microscope to capture high-resolution images with fine details of fabric surfaces. The CoMMonS dataset consists of 6,912 images covering 24 fabric samples in a controlled environment under varying imaging conditions such as lighting, zoom levels, geometric variations, and touching directions. We use this dataset both to assess the performance of existing deep learning-based algorithms and to develop our own method for material surface characterization in terms of fabric properties such as fiber length, surface smoothness, and toweling effect.

The main contributions of our research in this paper are summarized below:
\begin{enumerate}
    \item introducing a publicly available dataset for material surface characterization, which is the first of its kind to the best of our knowledge;
    \item formulating the understudied yet important problem of material surface characterization as a very fine-grained texture classification problem, and assessing the existing deep learning-based texture representation techniques for this task using our dataset;
    \item developing an innovative network architecture that integrates both low- and high-level CNN features to achieve a multi-level texture representation, maintaining both texture details and local spatial information, which outperforms the state-of-the-art deep texture representation techniques for both material surface characterization and texture/material recognition.
\end{enumerate}

% % chapter 2:
\section{Relevant Prior Works}
\label{sec:priorwork}

\subsection{Standard Pipeline for Texture Representation}

\begin{figure*}[htb]
\centering
{\includegraphics[width=15.5cm]{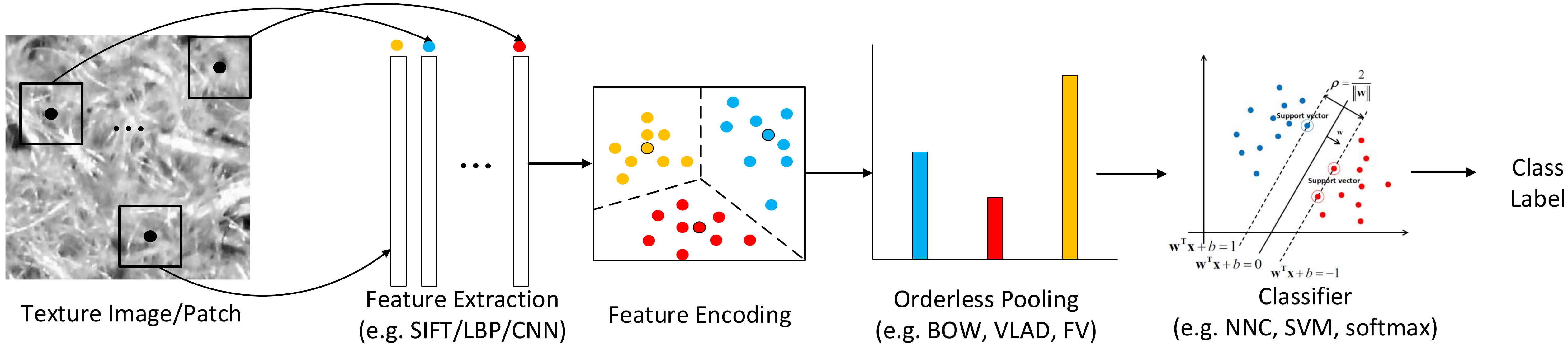}
\caption{The general BOW pipeline.}
\label{fig:General BOW pipeline}}
\end{figure*}

As reviewed in~\cite{cimpoi2015deep,liu2017local,liu2019bow}, texture representation techniques include filter-bank-based approaches, statistical models, bag of words (BOW) pipelines, and the latest CNN-based descriptors. Over the last decade, representation based on the BOW model, shown in Fig.~\ref{fig:General BOW pipeline}, has become a popular choice over others. BOW combined with local descriptors, such as the scale-invariant feature transform (SIFT)~\cite{lowe2004distinctive}, local binary pattern  (LBP)~\cite{ojala1996comparative}, or LBP variants~\cite{ hu2016completed, hu2017scale}, was the most widely used texture representation method. By assigning each local descriptor to its nearest visual word (i.e., a hard assignment), the BOW encoder calculates a histogram of visual word occurrences. To include richer information instead of simple occurrences, two popular extensions of BOW are vector of locally-aggregated descriptors (VLAD)~\cite{jegou2010aggregating} and Fisher vectors (FV)~\cite{perronnin2007fisher}. Different from BOW, VLAD accumulates the differences between a visual word and its corresponding local descriptors to aggregate first-order statistics of descriptors, while FV encodes both first- and second-order statistics of descriptors. With the advent of the deep learning technology, nowadays it is of interest to integrate these successful techniques into deep learning networks for more challenging real-world scenarios. 

\subsection{State-of-the-art Learning-based Texture Representation}

We mainly discuss two major learning-based methods: FV-CNN~\cite{cimpoi2015deep} and end-to-end learning~\cite{zhang2017deep,xue2018deep}. FV-CNN~\cite{cimpoi2015deep} computes FV~\cite{perronnin2007fisher} on top of generic descriptors such as SIFT, color features, or deep features such as pretrained deep convolutional activation features (DeCAF)~\cite{donahue2014decaf}. FV-CNN as a gold standard texture representation achieves state-of-the-art recognition results on texture datasets such as FMD~\cite{sharan2009material} and KTH-TIPS~\cite{hayman2004significance}. One shortcoming of the FV-CNN architecture is the separate learning of CNN feature extraction, texture encoding, and classifier training, which does not benefit from the labeled data. To jointly learn them together in an end-to-end manner, Zhang et al.~\cite{zhang2017deep} proposed a texture encoding layer, which builds the dictionary learning and feature pooling on top of the CNN architecture. This deep texture encoding network (Deep-TEN) learns an orderless representation, which performs well on texture or material recognition. But as textures do not always exhibit completely orderless patterns, local spatial information is still useful for differentiating them. To address this issue, Xue et al.~\cite{xue2018deep} presented a deep encoding pooling network (DEP), which fuses orderless texture encoding and local spatial information to yield enhanced performance over Deep-TEN. However, neither Deep-TEN nor DEP fully utilizes CNN features from different layers and resolutions, leaving room for further development.  

\subsection{Dataset Collection}

\begin{table*}[htb]
\begin{center}
\caption{Comparison between CoMMonS dataset and some publicly available texture and material datasets. The comparison follows the example in~\cite{liu2019bow} and reuses its relevant information.}
\label{tab: datasetcomparison}
\resizebox{1\textwidth}{!}{
    \begin{tabular}{|c|c|c|c|c|c|c|c|c|c|c|c|c|c|}
    \hline
         \multirow{2}{*}{Name} & Total  & \multirow{2}{*}{Classes}  &  Image & Gray/  & Imaging & Illumination     &  Rotation     & Viewpoint  & Scale       &  Image  &    \multirow{2}{*}{Scenes}   & Grain  & \multirow{2}{*}{Year}   \\
          & Images  &   &  Size & Color  &  Environment &  Changes    &   Changes    &  Changes &  Changes      &   Content &    & Level   &  \\
    % \hline
    %      Brodatz & 111  & 111  & 640$\times$640 &  Gray & Constrained  &    &     &     &      &  Objects  &    &  &  1966 \\
  \hline
     CUReT~\cite{dana1999reflectance}     & 5612  &  92 &  200$\times$200 & Color &  Constrained &  Yes  &  Small   &  Yes   &  No   &  Materials    & No  & Coarse & 1999  \\
    % \hline
    % Outex      & 8640   &  320 & 746$\times$538 & Color & Constrained  &  Yes  &  Yes   &     &      &  Materials &    &  &  2002 \\
    % \hline
    %  UIUC     & 1000  & 25  & 640\times640 & Gray  & Wild   &  Yes  & Yes    &    Yes & Yes     &  Materials &    &  &  2005 \\
    \hline
    KTHTIPS~\cite{caputo2005class}      &  810 & 10  & 200$\times$200 & Color & Constrained  & Yes   &  Small   &   Small  &   Yes   & Materials  & No     & Coarse &  2004 \\
    \hline
    % KTHTIPS2a      & 4608  & 11  & 200\times200& Color & Constrained  &   Yes   &  Small   &   Small  &   Yes     & Materials    & 2006   &  &   \\
    % \hline
    % KTHTIPS2b      & 4752  & 11  & 200\times200& Color & Constrained  &   Yes   &  Small   &   Small  &   Yes     & Materials    & 2006   &  &   \\
    % \hline
    % UMD      & 1000  &  25 & 1280\times960 & Gray & Wild   &    Yes &    Yes  &  Yes    &    Yes   & Objects    &   &  &  2009 \\
    % \hline
    % ALOT      & 25000  & 250  & 1536\times1024 & Color &    Constrained   & Yes    &     &      &   &  Materials   &  & &  2009 \\
    %   \hline
     FMD~\cite{sharan2009material}     & 1000  & 10  & 512$\times$384 & Color  & Wild   & Yes    & Yes    & No    & No     & Materials    &  In scene  & Coarse &  2009 \\
      \hline
     OpenSurfaces~\cite{bell2013opensurfaces}     &  10422  & 22  & Unfixed & Color &  Wild  & Yes   & Yes    &  Yes   &   Yes   & Materials    & In scene  & Coarse  & 2013  \\
    \hline
      DTD~\cite{cimpoi2014describing}    &  5640 & 47  & Unfixed & Color  & Wild   &Yes &    Yes &   No  & Yes     & Textures   &  In scene  & Fine  &  2014 \\
    \hline
      TUM Haptic Texture~\cite{strese2014haptic}    & 690 & 69  & 224$\times$ 224 & Color  & Controlled   &Yes &    Yes &   No  & Yes     & Materials   &   No  & Fine  &  2014 \\
    \hline
      MINC~\cite{bell2015material}    &  2996674 & 23  & Unfixed & Color & Wild  & Yes   & Yes    & Yes    &     Yes & Materials  & In scene & Coarse  & 2015  \\
      \hline
      MINC2500~\cite{bell2015material}    &  57500 & 23  & 362$\times$362 & Color & Wild  & Yes   & Yes    & Yes    &     Yes & Materials  &   In scene & Coarse &  2015 \\
      \hline
      4D Light-field~\cite{wang20164d}    &  30000 & 12  & 128$\times$ 128 & Color   & Constrained  &  No   & No    &  No    & No  & Materials   & In scene  & Coarse &   2016\\
      \hline   
     IC-CERTH Fabric~\cite{kampouris2016fine}    & 1266  &  9 & 640$\times$480 & Color  & Constrained  &   Yes  & No    &    No  & No  & Materials   & No  & Fine &    2016\\
      \hline 
      GTOS~\cite{xue2017differential}    & 34243  & 40  & 240$\times$240 & Color & Partially Constrained   & Yes   &Yes     &Yes     &   No   &  Materials  &  Scene  & Coarse & 2016  \\
    \hline 
      GTOS-mobile~\cite{xue2018deep}    & 34243  & 31  & 240$\times$240 & Color & Partially Constrained   & Yes   &Yes     &Yes     &  No    &  Materials  &  Scene  &  Coarse &  2018 \\
    \hline
    \textbf{CoMMonS (ours)}  & 6912  &  12 & 1920$\times$2560 & Color  &  Constrained &  Yes  &  Yes   &   No  &  Yes    & Materials  &   No    & Very fine & 2019 \\
    \hline

    \end{tabular}
}
\end{center}
\end{table*}

Over the years, texture/material datasets have transitioned from cropped stand-alone samples collected in lab settings (e.g. CUReT~\cite{dana1999reflectance},  KTH-TIPS\cite{caputo2005class}) to image sets collected in the wild (e.g. FMD~\cite{sharan2009material}, OpenSurfaces~\cite{bell2013opensurfaces}, LFMD~\cite{wang20164d}, and MINC~\cite{bell2015material}) with more diverse samples and real-world scene context. This transition led to explorations on how to generalize from one example of a material to another, such as real-world texture attributes studied in~\cite{cimpoi2014describing}.

Early studies of material surface analysis largely concentrated on reflectance-based in-lab constrained radiometric measurements~\cite{dana2016capturing} and the image-based modeling~\cite{xue2017differential}. The reflectance-based measurement captures intrinsic surface properties, while the image-based modeling captures surfaces with a single-view image in scene without the multiview reflectance information. Both modelings enable material recognition~\cite{cimpoi2015deep,bell2015material,liu2010exploring,zhang2015reflectance}, but the image-based modeling is more appropriate for real-world uses. Between the two modeling, Xue~\cite{xue2017differential} proposed a framework of augular differential imaging, which utilizes both rich radiometric cues and flexible image capture to enhance material recognition performance. Additionally, the tactile GelSight sensor~\cite{li2013sensing}, the phometric stero sensor~\cite{kampouris2016fine}, and the haptic acceleration signal~\cite{strese2014haptic} have all been explored. However, these imaging methods are not suitable for the material surface characterization problem under study in this paper, where fine details of material surface dictating the properties of interest can only be observed using microscopic imaging. A comparison between our CoMMonS dataset and existing public texture/material datasets is shown in Table~\ref{tab: datasetcomparison}.

% chapter 3: dataset description
\section{Dataset Description}
\label{sec:dataset}

\subsection{Image Acquisition System}

\begin{figure}[htb]
\centering
{\includegraphics[width=8.5cm]{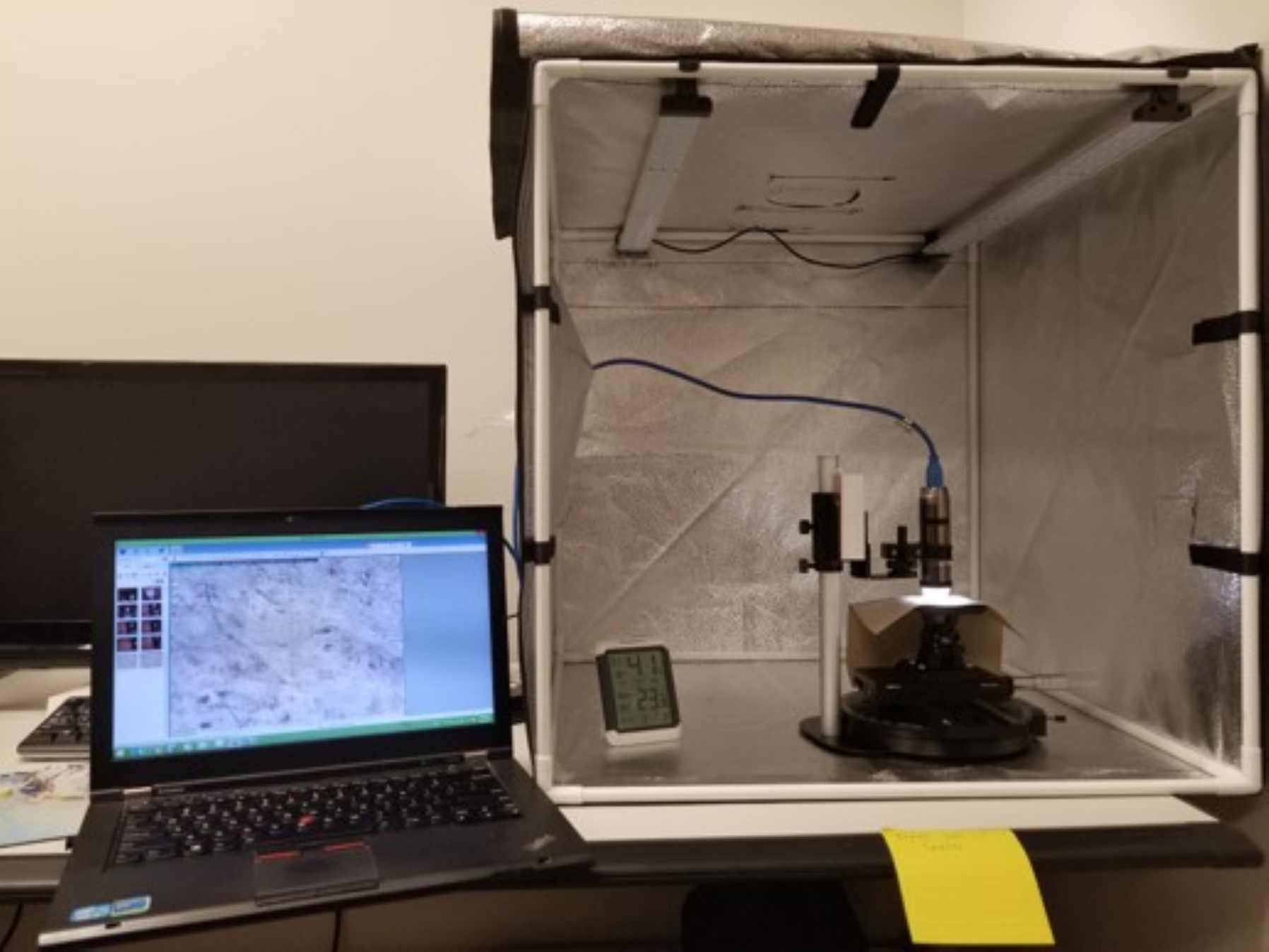}
\caption{An illustration of the data acquisition system. The microscope is mounted on a table stand and connected to a laptop. A fabric sample is placed right under the microscope and on top of a manual staging system. The system (except for the laptop) is set up within a photo light box to ensure controlled lighting conditions. Controllable lighting sources are available with both the light box and the microscope. A temperature/humidity monitor is placed inside the light box to keep record of the temperature and humidity while acquiring images.}
\label{fig:Dataset acquisition setup}}
\end{figure}   

\textbf{An imaging system}: A carefully designed imaging system is critical for the analysis of fabric surface images. This imaging system needs to be powerful enough to capture the fine details of fabric surfaces, which reveal the textural differences that characterize the fabrics. While a regular digital camera usually cannot accomplish this task, a commercial imager, Dino-lite AM73915MZT is appropriate for our imaging system. Some key features of this microscope include: an optical magnification power ranging from 10× to 220×; a high resolution of 2560×1920 pixels; an automatic magnification reading (AMR) function that enables automatic magnification rate recording; an extended depth of field (EDOF) that provides enhanced image quality at high magnification rates; and an enhanced dynamic range (EDR) that provides enhanced image quality for limited dynamic range conditions.

\textbf{A controlled environment}: The dataset should be a comprehensive collection, representing different types of fabrics imaged in various conditions. Such a comprehensive coverage is crucial for validating the system and the algorithms, ensuring a robust performance. Therefore, the data acquisition system needs to be established within a controlled environment. It has to include various environmental conditions as encountered in real-world settings, such as variations in lighting, zoom-in level of the camera lens, position of the fabric, etc. Our main focus was on establishing a staging system which is the most challenging component of such a controlled environment. Considering the large quantity of images to be acquired, our initial plan was to utilize commercial motorized staging systems for automated positioning during image acquisition, so that we could achieve accurate positioning, repeatability of experiments, efficient data acquisition, and significantly reduced acquisition time. However, because of the high cost of motorized staging systems, we switched to commercial manual staging systems that would satisfy our requirements in terms of accuracy, repeatability, and efficiency. The complete data acquisition system is shown in Fig.~\ref{fig:Dataset acquisition setup}.

\subsection{CoMMonS Dataset Collection}

\begin{figure*}[htb]
\centering
{\includegraphics[width=15cm]{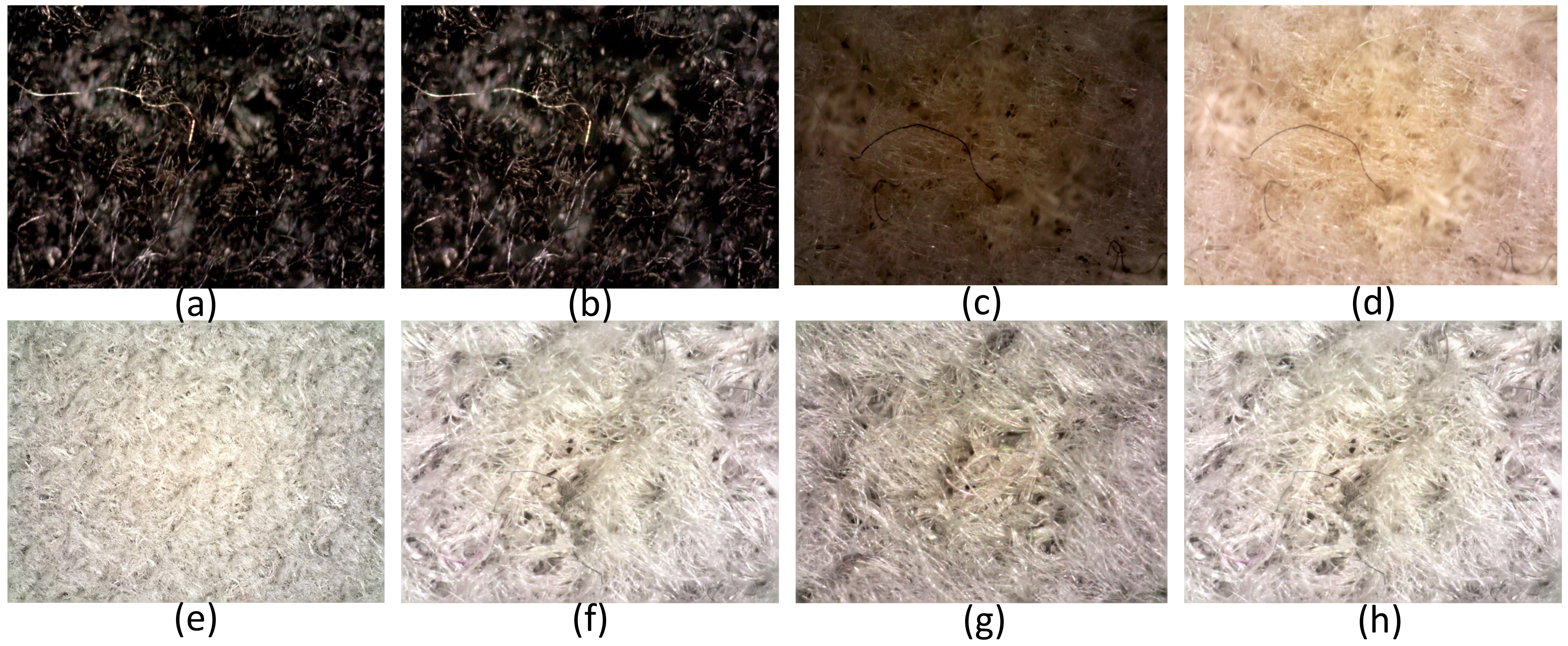}
\caption{Example images obtained with the data acquisition system: (a) sample ``S2'' without EDOF; (b) sample ``S2'' with EDOF, which significantly reduces the blurry areas; (c) sample ``S4'' without EDR; (d) sample ``S4'' with EDR, which enhances the image details not revealed clearly under the original condition; (e) sample ``S1'' with camera zoom level 50; (f) sample ``S1'' with camera zoom level 200; (g) sample ``S1'' along the ``pile'' pressing direction; and (h) sample ``S1'' along the opposite ``pile'' pressing direction.}
\label{fig:Dataset setting example}}
\end{figure*}

We created a comprehensive dataset of images captured at the fabric surfaces under varying conditions. We have 24 samples from ``S1'' to ``S24'' with subjective quality evaluation for three fabric properties, i.e., fiber length, smoothness and toweling effect, respectively. We acquired images for these samples under varying conditions including six translation positions, two rotation angles, two lighting conditions, two camera zoom levels, three camera function settings (Normal/EDOF/EDR), and two sample conditions regarding touching (or pressing) directions. The pressing directions are included because they will affect a human expert in evaluating the fabric properties. Combining all these conditions, we acquired a dataset of around 7000 images. Example images are shown in Fig.~\ref{fig:Dataset setting example}. We list key features of CoMMonS dataset and compare it with other counterparts in Table~\ref{tab: datasetcomparison}. 

\begin{figure}[htb]
\centering
{\includegraphics[width=9cm]{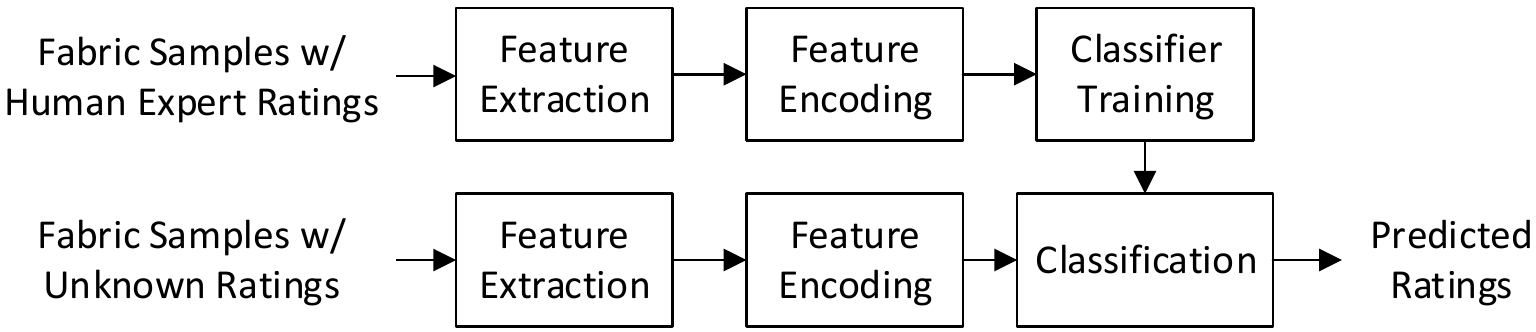}
\caption{Supervised Learning for material quality evaluation.}
\label{fig:supervisedlearningformaterialqualityevaluation}}
\end{figure}

Different from other datasets, our dataset focuses on material surface characterization for one material (fabric) in terms of one of three properties (fiber length, smoothness, and toweling effect), facilitating a very fine-grained texture classification. In this particular case, the dataset is used for a standard supervised problem of material quality evaluation, as shown in Fig.~\ref{fig:supervisedlearningformaterialqualityevaluation}. It takes fabric samples with human expert ratings as training inputs, and takes fabric samples without human subject ratings as testing inputs to predict quality ratings of the testing samples. The texture patches are classified into 4 classes according to each surface property measured by human sense of touch. For example, the human expert rates surface fiber length into 4 levels, from 1 (very short) to 4 (long), and similarly for smoothness and toweling effect. Because the samples all belong to the same type of fabric, the intra-class appearance variation is much smaller, making the classification much more challenging. Also, our images are of much higher resolution comparing to those from other datasets.

% % chapter 4: proposed method
\section{Proposed Learning-based Method for Material Surface Characterization}
\label{sec:proposedsystemmethod}

\begin{table*}[htb]
\centering
\caption{Architecture for adopting pretrained ResNet50.}
\resizebox{0.7\textwidth}{!}{
\begin{tabular}{|l|l|c|c|c|c|}
\hline
Modules & Layers                      &      Basic Blocks/Layers          & Ouput Size &  Multi-levels & LEM Output Size               \\
\hline
& Conv1                       & 7$\times$7, 64, stride 2                &       150$\times$150$\times$64           &  &     \\
\hline
\multirow{4}{*}{ResNet50} & Res1    & $\begin{bmatrix}
1\times 1,64  \\
3\times 3,64  \\
1\times 1,64
\end{bmatrix}\times 3$  &  75$\times$75$\times$64  & LEM1 & C=128  \\
\cline{2-6}
& Res2                        &      $\begin{bmatrix}
1\times 1,64  \\
3\times 3,64  \\
1\times 1,64
\end{bmatrix}\times 4$            &        38$\times$38$\times$128       &  LEM2  &      C=128           \\
\cline{2-6}
&  Res3                        & $\begin{bmatrix}
1\times 1,64  \\
3\times 3,64  \\
1\times 1,64
\end{bmatrix}\times 6$                &   14$\times$14$\times$256        &  LEM3 &     C=128                  \\
\cline{2-6}
& Res4                        & $\begin{bmatrix}
1\times 1,64  \\
3\times 3,64  \\
1\times 1,64
\end{bmatrix}\times 3$                    &    19$\times$19$\times$512 &  LEM4   &           C=128            \\
\hline
Classifier & FC          &       128$\times$4 = 512 =\textgreater{}n               &            n classes        & &     \\
\hline
\end{tabular}
}
\label{tab:proposedarchitecture}
\end{table*}

\begin{figure*}[htb]
% \begin{minipage}[b]{1.0\linewidth}
  \centering
  \centerline{\includegraphics[width=17.5cm]{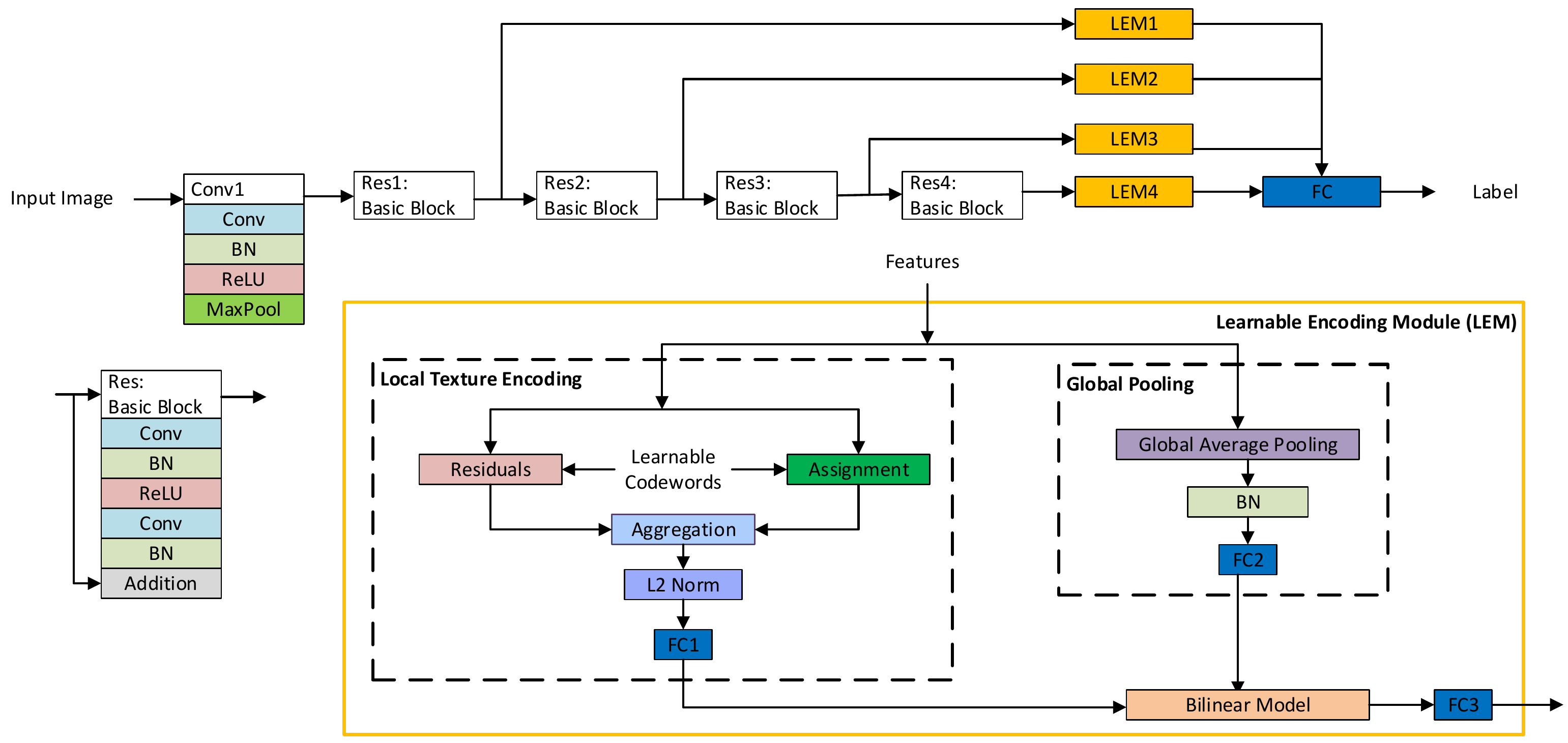}}
% \end{minipage}
\caption{Flowchart of our proposed method.}
\label{fig:flowchart}
\end{figure*}

To combine both low- and high-level CNN features, we propose a multi-level texture encoding and representation network (MuLTER) on top of our previous work~\cite{hu2019multi}, whose architecture is shown in Fig.~\ref{fig:flowchart} and Table~\ref{tab:proposedarchitecture}. We build the MuLTER on top of convolutional and non-linear layers pretrained on ImageNet~\cite{russakovsky2015imagenet} (e.g., ResNet18 or ResNet50~\cite{he2016deep}). In addition, we incorporate learnable encoding modules at each individual layer.

\subsection{Learnable Encoding Module (LEM)}
\label{subsec:deepten}

\begin{table}[htb]
\caption{Learnable Encoding Module (LEM). The 3rd colum shows the output sizes for an input image size of $224\times224\times3$ and the 4th column shows the basic blocks or layers used.}
\begin{tabular}{|l|l|c|c|}
\hline
Spatial & Layers                      & Output Size               & Basic Blocks/Layers                    \\
\hline
& Reshape                     & WH$\times$D                    & W$\times$H$\times$D =\textgreater{}WH$\times$D    \\

\hline
\multirow{2}{*}{Local} & Encoding                    & K$\times$D                     & K codewords                \\
\cline{2-4}

& Projection & 64 & FC1: KD =\textgreater{}64 \\
\hline

\multirow{2}{*}{Global} & Pooling                     & D                       & Average Pooling            \\
\cline{2-4}
& Projection                            & 64                        & FC2: 512 =\textgreater{} 64  \\
\cline{1-4}
L\&G & Bilinear              & 4096                      &        =\textgreater{}  $64^{2}$                   \\
\hline
& Projection                  & C=128                       & FC3: 4096 =\textgreater 128              \\

\hline
\end{tabular}
\label{tab:LEM}
\end{table}

For texture recognition in an end-to-end learning framework while maintaining texture details, the ``texture encoding'' layer was proposed~\cite{zhang2017deep}, which integrates dictionary learning and texture encoding in a single learnable model on top of convolutional layers, shown in Fig.~\ref{fig:flowchart}. It learns an inherent dictionary of local texture descriptors extracted from CNNs and generalizes robust residual encoders such as VLAD~\cite{jegou2010aggregating} and Fisher Vector~\cite{perronnin2010improving} through a ``residual'' layer calculated by pairwise difference between texture descriptors and the codewords of the dictionary. In ``assignment'' layer, assignment weights are calculated based on pairwise distance between texture descriptors and codewords and the ``aggregation'' layer converts the residuals vectors and the assignment weights into a full image representation. Thanks to the residual encoding, such image representations discarding frequently appearing features are helpful to domain transfer learning.

In addition to orderless texture details captured by the encoding layer, local spatial information are important visual cues, and the ``global pooling'' layer~\cite{xue2018deep} preserves local spatial information by average pooling. Then, a bilinear model~\cite{freeman1997learning} follows the texture encoding layer and the global pooling layer to jointly combine the two types of complementary information. We refer to the entire module as a learnable encoding module (LEM), shown in Fig.~\ref{fig:flowchart} and Table~\ref{tab:LEM}. Here we briefly introduce the notations. The input size of a LEM is $W\times H\times D$, where $W$, $H$, and $D$ denote the width, the height, and the feature channel dimension of the input volume, respectively. The codewords' number of the learnable dictionary is $K$.

\subsection{Multi-level Deep Feature Fusion}
\label{subsec:multi-levelfeaturefusion}

The multi-level feature fusion means the joint utilization of both low-level features and high-level features from Res1 to Res4 of ResNet50. ResNet50 uses 4 basic blocks of similar structures and one example of the basic block is shown in the left bottom of Fig.~\ref{fig:flowchart}. Given an input image with size $300\times300\times3$, after employing convolutional filters (i.e. Conv1, a default structure at the beginning of the Resnet family), the output size is $150\times150\times64$. Then we feed it into ResNet18. Here, we have four levels, Res1, Res2, Res3, and Res4.

The outputs from each level have different output sizes so we feed them into different sizes of LEMs. For example, for the first level, Res1 is followed by LEM1, where the output size of Res1 is $W\times H\times D=150\times 150\times 64$ and LEM1 converts it into a feature vector of dimension $C=128$. Whatever the input image size is, the same architecture shown in Table~\ref{tab:proposedarchitecture} can be used to produce a fixed-length (i.e., $C$) feature representation. Similar to the first level, we can repeat the procedure above to calculate a feature vector of dimension $C=128$ for level 2, 3, and 4. For local CNN-based texture descriptors at each level with either low-level features or high-level features, we preserve both texture details and local spatial information through their corresponding LEMs. To combine the features from different levels, we concatenate them and feed them into a classification layer. Assuming the number of classes is $n$, the classification layer maps the $4C$ feature vector to $n$ classes.

The multi-level architecture for texture encoding and representation has multiple advantages. First, the multi-level architecture makes it easy to adjust regarding which level of information should be fused. Second, it can be easily extended to other CNN models by adapting the size of LEMs and the number of levels. Third, all modules in the overall architecture are differentiable, so the network can be trained with back propagation in an end-to-end texture encoding and representation network. Last but not the least, this architecture produces a compact yet discriminative representation with a full image representation with a dimension of 512.

% % chapter 5: experiments
\section{Experimental Results}
\label{sec:experimentalresults}

\subsection{Implementation and Evaluation on CoMMonS}

\begin{figure*}[htb]
\centering
{\includegraphics[width=15.5cm]{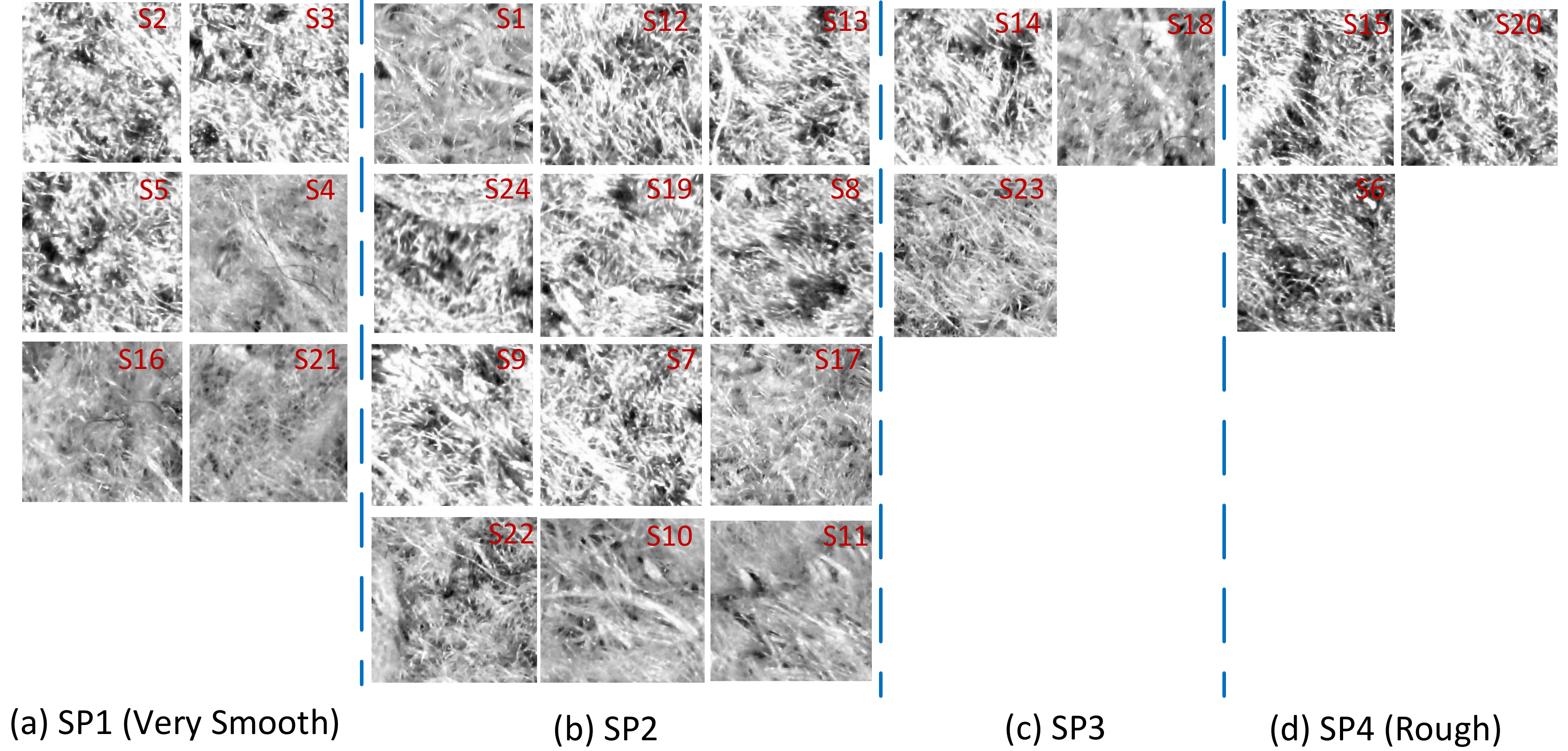}
\caption{Patch examples regarding the ``smoothness'' property.}
\label{fig: patch sample example}}
\end{figure*}

\textbf{Data Preparation}: We performed six-fold cross validation for training and testing splits. In our work, images were acquired at six non-overlapping locations of each fabric sample. Taking one fold as an example, patches of size $300\times300$ extracted from images of location 2, location 3, $\cdots$, and location 6 are used to form the training set, which is used to train and learn the feature encoder and the classifier. Then, patches of the same size extracted from images of location 1 of all fabric samples are used to form the testing set, which is used to test how good the feature encoder and the classifier are for predicting the characteristics of a given patch. Patch examples of different ``smoothness'' levels from ``very smooth'' to ``rough'' are shown in Fig.~\ref{fig: patch sample example}. We repeat this procedure for other five data splits. Such six-fold cross validation is used for texture classification evaluation of different representation algorithms later. The training set is used to train FV-CNN or end-to-end learning approaches and the testing dataset is used to test how good they are for predicting the characteristics of a given patch.

\textbf{Performance Evaluation}: We run experiments on a PC (Nvidia GeForce GTX1070, RAM: 8GB) and evaluate state-of-the-art methods including FV-CNN~\cite{donahue2014decaf} and DEP~\cite{xue2018deep}. The experimental settings for DEP~\cite{xue2018deep} and MuLTER are: learning rate starting at 0.1 and decaying every 10 epochs (step = 10) by a factor of 0.1, batch size 16, and the total number of epochs 30. The number of codewords $K$ is set to 32. For a fair comparison to our MuLTER method using fully connected layer and softmax as a classifier, we extract FV-CNN~\cite{donahue2014decaf} features combining pre-trained CNN features and feature encoding and apply the same classifier and the same setting (i.e. staring learning rate = 0.1, step = 10, epochs = 30, and batch size =16). Regarding FV-CNN, we evaluate two pretrained models, VGG-M and VGG-VD. The extensive comparisons of the four methods for each surface property and two zooming levels are shown in Table~\ref{tab: comparasionresultsonCoMMonSforfiberlength_z50} to Table~\ref{tab: comparasionresultsonCoMMonSforTowelling_z200}, respectively. 

\begin{table}[htb]
\begin{center}
\caption{Comparison with state-of-the-art algorithms in $\%$ on the CoMMonS dataset for the ``Fiber length'' property. (zoom: 50, rotation: -30$^\circ$)}
\label{tab: comparasionresultsonCoMMonSforfiberlength_z50}
\resizebox{0.5\textwidth}{!}{
    \begin{tabular}{|c|c|c|c|c|}
    \hline
    Data     &  FV-CNN~\cite{donahue2014decaf} &  FV-CNN~\cite{donahue2014decaf}&
    \multirow{2}{*}{DEP~\cite{xue2018deep}} & \multirow{2}{*}{\textbf{Proposed}} \\
    Splits & (VGG-M) & (VGG-VD)& & \\
    \hline
    Split 1   & 70.5 &  67.4 &  58.4  & 65.3\\
    \hline
    Split 2  & 52.4  & 58.9  & 58.2  & 58.7 \\
    \hline
    Split 3  & 64.2  &  55.3 & 50.3 & 52.1 \\
    \hline 
    Split 4  & 59.2  &  55.8 & 67.6 & 72.4 \\
    \hline
    Split 5  & 58.2  & 59.2  & 63.2 & 66.6 \\
    \hline
    Split 6  & 63.4  & 54.2  & 44.2 & 57.1\\
    \hline
    Average  & 61.3$\pm$5.6 & 58.5$\pm$4.4 & 57.0$\pm$7.8 & \textbf{62.0$\pm$6.7} \\
    % \hline
    % Difference  & 0.7 & 3.5 & 5.0 & 0 \\
    \hline
    \end{tabular}
}
\end{center}
\end{table}

\begin{table}[htb]
\begin{center}
\caption{Comparison with state-of-the-art algorithms in $\%$ on the CoMMonS dataset for the ``Smoothness'' property. (zoom: 50, rotation: -30$^\circ$)}
\label{tab: comparasionresultsonCoMMonSforsmoothness_z50}
\resizebox{0.5\textwidth}{!}{
    \begin{tabular}{|c|c|c|c|c|}
    \hline
    Data     &  FV-CNN~\cite{donahue2014decaf} &  FV-CNN~\cite{donahue2014decaf}&
    \multirow{2}{*}{DEP~\cite{xue2018deep}} & \multirow{2}{*}{\textbf{Proposed}} \\
    Splits & (VGG-M) & (VGG-VD)& & \\
    \hline
    Split 1   & 52.5  &   44.8 & 54.8   & 55.6 \\
    \hline
    Split 2  &  55.7  &  60.8 & 57.5  & 65.6  \\
    \hline
    Split 3  &  59.4  & 56.0  & 57.3 & 60.8 \\
    \hline
    Split 4   &  56.9   &  53.3   & 51.9 & 54.4  \\
    \hline
    Split 5  & 55.6   &  52.9  & 54.2 & 58.1  \\
    \hline
    Split 6  &  56.3   &  50.2  & 65.6 & 59.6 \\
    \hline
    Average   & 56.1$\pm$2.0  & 53.0$\pm$4.9  & 56.9$\pm$4.3 & \textbf{59.0$\pm$3.7}  \\
    \hline
    \end{tabular}
}
\end{center}
\end{table}

\begin{table}[htb]
\begin{center}
\caption{Comparison with state-of-the-art algorithms on the CoMMonS dataset regarding the ``Toweling'' property. (zoom: 50, rotation: -30$^\circ$)}
\label{tab: comparasionresultsonCoMMonSforTowelling_z50}
\resizebox{0.5\textwidth}{!}{
    \begin{tabular}{|c|c|c|c|c|}
    \hline
    Data     &  FV-CNN~\cite{donahue2014decaf} &  FV-CNN~\cite{donahue2014decaf}&
    \multirow{2}{*}{DEP~\cite{xue2018deep}} & \multirow{2}{*}{\textbf{Proposed}} \\
    Splits & (VGG-M) & (VGG-VD)& & \\
    \hline
    Split 1   &  58.4  &  51.6  & 51.6   & 61.6 \\
    \hline
    Split 2   &   57.8 & 53.4 & 55.3 & 57.5 \\
    \hline
    Split 3  &  55.9  & 51.9 & 47.5 & 55.6 \\
    \hline
    Split 4   &  48.1  & 57.2 & 41.9 & 48.8 \\
    \hline
    Split 5   &  56.3  & 51.0 & 52.2 & 56.3 \\
    \hline
    Split 6  & 60.0 & 62.8 & 60.0 & 57.8 \\
    \hline
    Average   & 56.1$\pm$3.8   & 54.7$\pm$4.2 & 51.4$\pm$5.7 & \textbf{56.3$\pm$3.8} \\
    \hline
    \end{tabular}
}
\end{center}
\end{table}

\begin{table}[htb]
\begin{center}
\caption{Comparison with state-of-the-art algorithms in $\%$ on the CoMMonS dataset for the ``Fiber length'' property. (zoom: 200, rotation: -30$^\circ$)}
\label{tab: comparasionresultsonCoMMonSforfiberlength_z200}
\resizebox{0.5\textwidth}{!}{
    \begin{tabular}{|c|c|c|c|c|}
    \hline
    Data     &  FV-CNN~\cite{donahue2014decaf} &  FV-CNN~\cite{donahue2014decaf}&
    \multirow{2}{*}{DEP~\cite{xue2018deep}} & \multirow{2}{*}{\textbf{Proposed}} \\
    Splits & (VGG-M) & (VGG-VD)& & \\
    \hline
    Split 1  &  42.1   & 45.5  &  41.8 & 46.3\\
    \hline
    Split 2  & 46.1    & 46.3 & 53.2  & 51.8\\
    \hline
    Split 3  & 53.7   & 52.6 & 57.7 &  58.2\\
    \hline
    Split 4  & 44.2    & 52.9 & 48.9   &  63.9\\
    \hline
    Split 5  &  43.2   & 51.3 & 47.6 &  52.4\\
    \hline
    Split 6  &   53.4   &  60.0  &  51.3  & 55.0\\
    \hline
    Average  &  47.1$\pm$4.7 &  51.4$\pm$4.8  &  50.1$\pm$4.9   & \textbf{54.6$\pm$5.5} \\
    \hline
    \end{tabular}
}
\end{center}
\end{table}

\begin{table}[htb]
\begin{center}
\caption{Comparison with state-of-the-art algorithms in $\%$ on the CoMMonS dataset for the ``Smoothness'' property. (zoom: 200, rotation: -30$^\circ$)}
\label{tab: comparasionresultsonCoMMonSforsmoothness_z200}
\resizebox{0.5\textwidth}{!}{
    \begin{tabular}{|c|c|c|c|c|}
    \hline
    Data     &  FV-CNN~\cite{donahue2014decaf} &  FV-CNN~\cite{donahue2014decaf}&
    \multirow{2}{*}{DEP~\cite{xue2018deep}} & \multirow{2}{*}{\textbf{Proposed}} \\
    Splits & (VGG-M) & (VGG-VD)& & \\
    \hline
    Split 1  & 50.4  & 47.3   & 51.9  & 51.5\\
    \hline
    Split 2  &  42.5 & 40.0    &  46.5 & 46.0\\
    \hline
    Split 3  & 50.8 &   44.0  & 56.9  &  52.3\\
    \hline
    Split 4  &  46.0  &  50.2    &  51.5  & 55.2 \\
    \hline
    Split 5  & 41.7   &   45.6   & 48.8  & 49.0 \\
    \hline
    Split 6  & 51.3  &   43.3   &  48.8   & 52.9\\
    \hline
    Average  & 47.1$\pm$ 4.0  & 45.1$\pm$3.2  & 50.7$\pm$3.3    & \textbf{51.2$\pm$2.9} \\
    \hline
    \end{tabular}
}
\end{center}
\end{table}

\begin{table}[htb]
\begin{center}
\caption{Comparison with state-of-the-art algorithms on the CoMMonS dataset for the ``Toweling'' property. (zoom: 200, rotation: -30$^\circ$)}
\label{tab: comparasionresultsonCoMMonSforTowelling_z200}
\resizebox{0.5\textwidth}{!}{
    \begin{tabular}{|c|c|c|c|c|}
    \hline
    Data     &  FV-CNN~\cite{donahue2014decaf} &  FV-CNN~\cite{donahue2014decaf}&
    \multirow{2}{*}{DEP~\cite{xue2018deep}} & \multirow{2}{*}{\textbf{Proposed}} \\
    Splits & (VGG-M) & (VGG-VD)& & \\
    \hline
    Split 1  &  39.1  &   47.8    & 39.4  & 47.5  \\
    \hline
    Split 2  & 50.6  &   45.6   & 41.9  & 47.5 \\
    \hline
    Split 3  &  50.3  &  50.9   & 40.9  & 42.8  \\
    \hline
    Split 4  & 37.5  &   38.4    &  49.4  & 47.2  \\
    \hline
    Split 5  & 38.8  &  43.8    & 45.9  & 44.7\\
    \hline
    Split 6  & 49.7 &  46.9   & 53.1    & 53.8\\
    \hline
    Average  & 44.3$\pm$5.9 &   45.6$\pm$3.9  &     45.1$\pm$4.9 & \textbf{47.3$\pm$3.4} \\
    \hline
    \end{tabular}
}
\end{center}
\end{table}

Taking the comparison regarding the ``Smoothness'' property under the zooming level ``50'' as an example, we discuss the performance of different methods on fabric surface property characterization. The classification accuracy of all six splits and their average are shown in Table~\ref{tab: comparasionresultsonCoMMonSforsmoothness_z50}. The best feature extraction algorithm (the one we proposed) achieved an average accuracy of $59.0\%$, which is $2.1\%$ higher over its closest counterpart, DEP~\cite{xue2018deep}, $2.9\%$ better than FV-CNN(VGG-M)~\cite{donahue2014decaf}, and $6.0\%$ more accurate than FV-CNN(VGG-VD)~\cite{donahue2014decaf}. From Table~\ref{tab: comparasionresultsonCoMMonSforfiberlength_z50} to Table~\ref{tab: comparasionresultsonCoMMonSforTowelling_z200}, we highlight the highest accuracy in bold. Our proposed method, MuLTER, achieved the best in all six tables, while the other methods in comparison did not perform as consistently. The improvement over the second best method is as follows: $0.7\%$ in Table~\ref{tab: comparasionresultsonCoMMonSforfiberlength_z50}, $2.1\%$ in Table~\ref{tab: comparasionresultsonCoMMonSforsmoothness_z50}, $0.2\%$ in Table~\ref{tab: comparasionresultsonCoMMonSforTowelling_z50}, $3.2\%$ in Table~\ref{tab: comparasionresultsonCoMMonSforfiberlength_z200}, $0.5\%$ in Table~\ref{tab: comparasionresultsonCoMMonSforsmoothness_z200}, and $1.7\%$ in Table~\ref{tab: comparasionresultsonCoMMonSforTowelling_z200}. 
This supports our claim that the end-to-end texture representation with multi-level feature fusion is capable of capturing the unique characteristics of fabrics in terms of fabric length, smoothness, and toweling effect better than its state-of-the-art counterparts. Although the overall accuracy rates are not high across the experiments, they are reasonable considering the challenging nature of the dataset. Interestingly, for all experiments, the performance is typically higher for fiber length, followed by smoothness, and lowest for toweling effect. We believe this is because fiber length is the most well-defined and most apparent among the three properties. On the contrary, toweling effect is an irregularity condition of the fabric, which is typically distributed very sparsely among the data samples. Additionally, comparing the classification accuracy of the same property under different zooming levels, we observe that patches with zooming level ``200'' are more difficult to be differentiated than those of zooming level ``50''. We believe this performance degradation is caused by a lack of necessary global or macro information. Therefore, capturing fine details with large zooming levels and maintaining global/macro information is a trade-off that needs to be carefully considered when designing the data collection protocol.

\subsection{Evaluation on Public Standard Datasets}
\textbf{Dataset Details}: To show the performance of our proposed method as a general texture representation technique for texture/material recognition, we test it on two recent challenging texture/material datasets: materials in context database (MINC)-2500~\cite{bell2015material} and ground terrain database (GTOS)-mobile~\cite{xue2018deep}. The MINC dataset is an order of magnitude larger than previous texture and material datasets (such as KTH-TIPS~\cite{caputo2005class} and FMD~\cite{sharan2009material}), while being more diverse and well-sampled across its 23 categories. For a fair comparison with other methods, we use MINC-2500 (i.e. a subset of MINC with 2500 patches per category). GTOS-mobile is a dataset including images for ground terrain regions captured by mobile phones. It consists of 31 classes such as grass, brick, soil, etc., and can be used for material classification. The GTOS-mobile is challenging because of its realistic capturing conditions (i.e. a mobile imaging device, hand-held video, and uncalibrated capture). Compared with GTOS-mobile, MINC-2500 is a more general one.

\textbf{Training Procedure}:  Following the standard testing protocol of MINC-2500 and GTOS-mobile, we use the same data argumentation and training procedure. We resize images to $256\times 256$ and randomly crop patches to $224\times224$. For the training part, we augment data using horizontal flips with a 50$\%$ probability. For a fair comparison with~\cite{xue2018deep}, we build a ResNet18 for the GTOS-mobile dataset and build a ResNet50 for the MINC-2500 dataset. As mentioned in Sec.~\ref{subsec:multi-levelfeaturefusion}, our method is easily extended to other CNN models (e.g. ResNet50) by adapting the size of LEMs. Our experimental settings are: learning rate starting at 0.01 and decaying every 10 epochs by a factor of 0.1, batch size 128 for GTOS-mobile and 32 for MINC-2500, momentum 0.9, and the total number of epochs 30. The number of codewords $K$ is set to 8 for GTOS-mobile and 32 for MINC-2500. The result is shown in Table~\ref{tab: comparasionresultsonGTOSandminc}, which shows the superior recognition accuracy of our proposed multi-level architecture. We run experiments on a PC (Nvidia GeForce GTX1070, RAM: 8GB).

\textbf{Performance Evaluation}: 
We evaluated our method and compared with other state-of-the-art methods on these two datasets, as shown in Table~\ref{tab: comparasionresultsonGTOSandminc}. The results for ResNet~\cite{he2016deep}, FV-CNN~\cite{cimpoi2014describing}, and Deep-TEN~\cite{zhang2017deep} were borrowed from~\cite{xue2018deep}. The results for DEP were generated using codes~\cite{dep_github} provided by the authors. On the MINC-2500 dataset, our method achieved a recognition accuracy of 82.2\%, which outperforms Deep-TEN by 1.8$\%$ and DEP by 1.2$\%$. On the GTOS-mobile dataset, the recognition accuracy of our method is 78.2$\%$, which is 4.0$\%$ better than Deep-TEN and 1.2$\%$ better than DEP. The reason behind our enhanced performance is that our method fuses multi-level CNN features in a distinctive and compact way while Deep-TEN and DEP only use features from a single level.

\begin{table}[t]
\begin{center}
\caption{Comparison with state-of-the-art algorithms on the MINC-2500 and the GTOS-mobile datasets in $\%$.}
\label{tab: comparasionresultsonGTOSandminc}
\resizebox{0.42\textwidth}{!}{
    \begin{tabular}{|c|c|c|c|}
    \hline
    Methods     &  MINC-2500~\cite{bell2015material} & GTOS-mobile~\cite{xue2018deep} \\
    \hline
    ResNet~\cite{cimpoi2015deep}    & N/A        &     70.8 \\
    %~\cite{xue2018deep}    \\
    \hline
    FV-CNN~\cite{cimpoi2015deep}      & 63.1        &     N/A    \\         \hline %~\cite{xue2018deep}        &     NA    \\         \hline
    Deep-TEN~\cite{zhang2017deep}  & 80.4 %~\cite{xue2018deep} %81.5\% (*)
    & 74.2%~\cite{xue2018deep} %77.6\% (*)
    \\
    \hline
    DEP~\cite{xue2018deep}  &  81.0  %(*)
    & 77.0 %~\cite{xue2018deep}%74.8\% (*)
    \\
    \hline
    \textbf{Proposed}  & \textbf{82.2} %(*)
    & \textbf{78.2} %(*)
    \\
    \hline
    \end{tabular}
}
\end{center}
\end{table}

% chapter 6: conclusion
\section{Conclusion}
\label{sec:conclusion}

In this paper, we formulated the problem of characterizing a material in terms of a certain property as a very fine-grained texture classification problem. Intrinsically, this is more challenging than merely recognizing a material, mainly because of the much smaller inter-class appearance variations and the fact that some material properties are more latent than apparent. We have successfully developed an objective vision-based system to assess fabric hand thereby laying the foundation for a “real-time” quality assessment system. Using microscopy imaging, we created a dataset of fabric surfaces, CoMMonS, a first-of-its-kind public dataset geared toward this understudied problem. We assessed the state-of-the-art deep learning-based texture representation techniques using CoMMonS, and demonstrated that they are inadequate for such surface characterization tasks. In addition, we proposed an innovative deep learning network architecture, MuLTER, that extracts both low-level and high-level CNN features to achieve a multi-level texture representation. With MuLTER, we were able to achieve enhanced performance over the state-of-the-art deep learning algorithms not only for material surface characterization using CoMMonS, but also for more general texture/material recognition. This proved the value of integrating CNN features from multiple levels with the embedment of a learnable encoding module at each level for texture representation. Our exploration here provided a unique benchmark for material surface characterization and very fine-grained texture classification, which will hopefully be useful for a wide rage of real-world application scenarios.

\section*{Acknowledgment}
The work was partially funded by Kolon Industries, Inc. through the Kolon Center for Lifestyle Innovation at Georgia Tech.

\ifCLASSOPTIONcaptionsoff
  \newpage
\fi



\bibliographystyle{IEEEtran}

\clearpage
\newpage

\bibliography{refs}


% \end{thebibliography}


% \end{IEEEbiography}


% % insert where needed to balance the two columns on the last page with
% %\newpage



%\vfill


\end{document}